\documentclass[journal=jpccck,manuscript=article]{achemso}

\usepackage[version=3]{mhchem} 
\setkeys{acs}{maxauthors = 10, etalmode=truncate}
\usepackage[dvipsnames]{xcolor}

\author{Robert J. Kirby}
\affiliation{Department of Chemistry, Princeton University, Princeton, New Jersey 08544, USA}
\author{Angela Montanaro}
\affiliation{Department of Physics, Università degli Studi di Trieste, Trieste I-34127, Italy}
\alsoaffiliation{Elettra Sincrotrone Trieste S.C.p.A., Basovizza, Trieste I-34012, Italy}
\author{Francesca Giusti}
\affiliation{Department of Physics, Università degli Studi di Trieste, Trieste I-34127, Italy}
\alsoaffiliation{Elettra Sincrotrone Trieste S.C.p.A., Basovizza, Trieste I-34012, Italy}
\author{Andr\'e Koch-Liston}
\affiliation{Department of Chemistry, Princeton University, Princeton, New Jersey 08544, USA}
\author{Shiming Lei}
\affiliation{Department of Chemistry, Princeton University, Princeton, New Jersey 08544, USA}
\author{Ioannis Petrides}
\affiliation{College of Letters and Science, UCLA, Los Angeles, CA 90095 USA}
\author{Prineha Narang}
\affiliation{College of Letters and Science, UCLA, Los Angeles, CA 90095 USA}
\author{Kenneth S. Burch}
\affiliation{Department of Physics, Boston College, Chestnut Hill, Massachusetts 02467, USA}
\author{Daniele Fausti}
\affiliation{Lehrstuhl f{\"u}r Festk{\"o}rperphysik, Friedrich-Alexander-Universit{\"a}t Erlangen-Nürnberg, Erlangen 91058, Germany}
\alsoaffiliation{Department of Physics, Università degli Studi di Trieste, Trieste I-34127, Italy}
\alsoaffiliation{Elettra Sincrotrone Trieste S.C.p.A., Basovizza, Trieste I-34012, Italy}
\author{Gregory D. Scholes}
\affiliation{Department of Chemistry, Princeton University, Princeton, New Jersey 08544, USA}
\author{Leslie M. Schoop}
\email{lschoop@princeton.edu}
\affiliation{Department of Chemistry, Princeton University, Princeton, New Jersey 08544, USA}

\title{Ultrafast Dynamics of the Topological Semimetal GdSb$_{x}$Te$_{2-x-\delta}$ In the Presence and Absence of a Charge Density Wave}

\abbreviations{IR,NMR,UV}

\begin{document}

\begin{tocentry}

\includegraphics{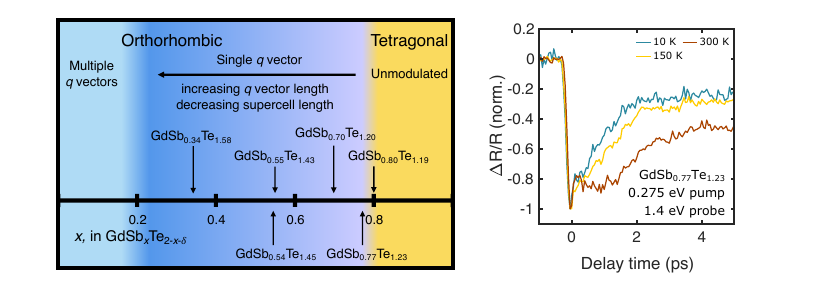}





\end{tocentry}

\begin{abstract}

Time-resolved dynamics in charge-density-wave materials have revealed interesting out-of-equilibrium electronic responses. However these are typically only performed in a single material possessing a CDW. As such, it is challenging to separate subtle effects originating from the CDW. Here, we report on the ultrafast dynamics of the GdSb$_{x}$Te$_{2-x-\delta}$ series of materials where E$_{F}$ can be tuned, resulting in a change from an undistorted tetraganal phase to a CDW with a wavevector that depends on $x$. Using mid-infrared, near-infrared, and visible excitation, we find the dynamics are sensitive to both E$_{F}$ and the presence of the CDW. Specifically, as the Sb content of the compounds increases, transient spectral features shift to higher probe energies. In addition, we observe an enhanced lifetime and change in the sign of the transient signal upon removing the CDW with high Sb concentrations. Finally, we reveal fluence- and temperature-dependent photo-induced responses of the differential reflectivity, which provide evidence of transient charge density wave suppression in related telluride materials. Taken together our results provide a blueprint for future ultrafast studies of CDW systems.

\end{abstract}


\section{Introduction}

Materials with highly lower-dimensional band structures are frequently known to undergo charge density wave (CDW) formation\cite{gruner1988dynamics}, in which the electron density and crystal lattice become spatially modulated. CDW materials are commonly studied for this electron-lattice correlation, which often occurs at more easily attainable conditions than similar phenomena in other correlated materials, like superconductors. The out-of-equilibrium response of the CDW state has also been extensively studied through, e.g., excitation with ultrashort pulsed lasers to understand how the various competing degrees of freedom react and subsequently interact as they return to equilibrium\cite{demsar1999single, demsar2002femtosecond, tomeljak2009dynamics, hellmann2010ultrafast, mohr2011nonthermal, lin2020optical, anikin2020ultrafast, monney2016revealing}.  ``Hidden'' metastable CDW states which are inaccessible via equilibrium routes have also been found\cite{stojchevska2014ultrafast, sun2018hidden, yoshikawa2021ultrafast, maklar2022coherent}. However, to date, these studies have typically been limited to a single compound and thus cannot uniquely separate CDW signatures.

Many of the early studies on the interaction of CDWs and light focused on blue bronzes and other oxide materials\cite{sagar2007coherent, sagar2008raman, demsar1999single, tomeljak2009dynamics, ren2004ultrafast}; recently, however, rare earth tritelluride compounds have been hotly investigated\cite{sacchetti2006chemical, yusupov2008single, schmitt2008transient, hu2011, hu2011optical, lavagnini2012infrared, hu2014coexistence, chen2014revealing, rettig2016persistent, chen2019raman, zong2019dynamical, zong2019evidence, kogar2020light, wang2022axial}. For example, the axial Higgs mode \textemdash\ evidence of an unconventional CDW due to multiple symmetries breaking \textemdash\ was recently observed at 300 K in GdTe$_{3}$ through quantum interference pathways in Raman scattering experiments\cite{wang2022axial}. One possible rationale behind these materials' popularity is that they allow a systematic tuning of the system, as many of them, from La to Tm (except Pm and Eu), exhibit at least one CDW\cite{malliakas2006divergence, ru2008effect, hu2014coexistence}; Tb \textendash\ Tm host two CDWs with different critical temperatures, T$_{C}$\cite{ru2008effect, banerjee2013charge}. In addition, a clear trend of the CDW common to all of the compounds is observed, where the critical temperature decreases with increasing atomic number; on the contrary, the T$_{C}$ of the second CDW increases with atomic number\cite{ru2008effect}.  Photoexcitation is hypothesized to enhance the CDW nesting conditions in DyTe$_{3}$\cite{rettig2016persistent}. In LaTe$_3$, however, ultrafast laser excitation can transiently suppress or even completely ``melt" the $c$-direction CDW, depending on the fluence\cite{zong2019evidence, zong2019dynamical}. A suite of time-resolved electronic and structural probes were then used to show that the recondensation of the CDW was hindered by topological defects in the material which slowed the return of phase coherence. LaTe$_{3}$ was also observed to host a photo-induced non-equilibrium incommensurate CDW along the $a$-axis, orthogonal to the equilibrium CDW\cite{kogar2020light}. 
Despite this, no studies have addressed what happens to the dynamics when the Fermi level is tuned or the CDW is completely removed.

Photoexcitation is usually thought to quench the CDW amplitude, but this is not always the case. While excitation with high energy light can often be described by a sudden increase of electronic temperature, hence, quenching the CDW order parameter \textemdash\ excitation with low energy light can give a different response and enhanced ``order'' \cite{rettig2016persistent} or metastable phases can be triggered. Having a system in which the Fermi surface nesting conditions and CDW gap can be continuously tuned (e.g., with electron filling) is of paramount importance when studying the response of CDWs to photoexcitation. Such a system can, for example, be capable of distinguishing dynamics associated with sudden increases in electronic temperatures due to other collective coherent responses such as changes to the nesting conditions\cite{rettig2016persistent}, reduced screening\cite{montanaro2022anomalous}, or coherent lattice dynamics\cite{kenji1998coherent, rettig2014coherent}. GdSb$_{x}$Te$_{2-x-\delta}$ ($\delta$ represents a vacancy concentration, see Fig. 1), in which the nesting conditions and CDW q vectors can be continuously tuned with electron filling (e.g., with Sb doping), provides one such rare opportunity\cite{lei2019charge, lei2021band}.

\begin{figure*}
\includegraphics[width=1\textwidth]{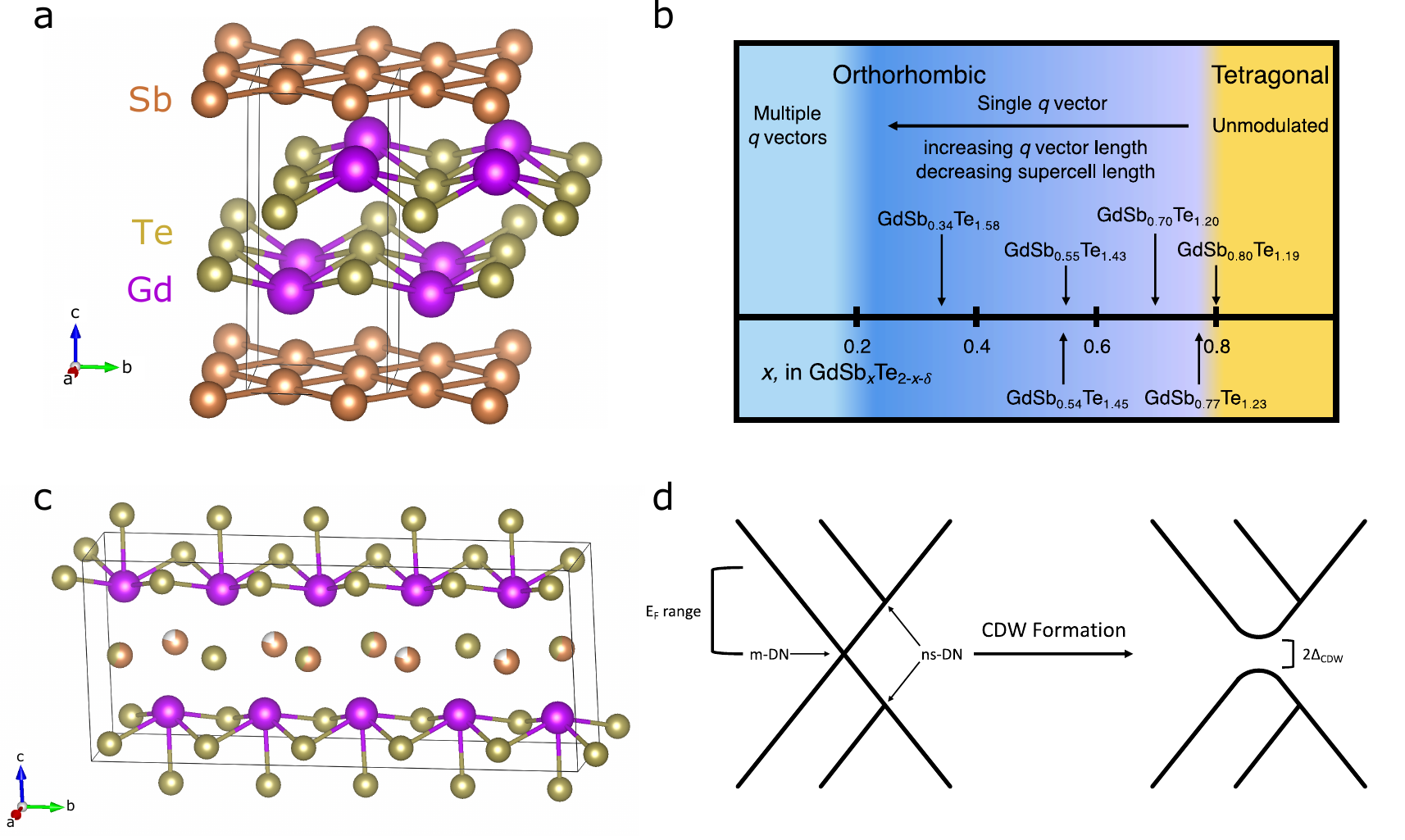}
\caption{\label{GdSbTe_struct_cdw} \textbf{a} Crystal structure of tetragonal GdSbTe. As more Te is incorporated, it replaces Sb in the square net and the unit cell becomes orthorhombic. This is also accompanied by supercell formation dependent on the Sb content. \textbf{b} Room temperature phase diagram of the GdSb$_{x}$Te$_{2-x-\delta}$ system as a function of Sb content, $x$\cite{lei2019charge}. As $x$ decreases from 1, the material enters the orthorhombic CDW phase between $x=0.80$ and $x=0.77$. This new CDW phase consists of a single q vector that increases in magnitude with decreasing $x$. Around $x=0.20$, the material begins to host more complex distortions with multiple q vectors. The formulae included in the diagram are studied in this work; compounds above and below the $x$-axis are probed in the near-infrared and visible regions, respectively.  \textbf{c} An example of a five-fold supercellular structure for an orthorhombic compound with $x \approx$ 0.5. Note the introduction of Te as well as vacancies in the Sb square net. \textbf{d} A general schematic of the electronic structure of these materials that highlights the impact of CDW formation, that is, that Dirac nodes that arise from mirror symmetry (m-DN) are gapped by some amount $2\Delta_{CDW}$ while nodes that arise from non-symmorphic symmetry (ns-DN) remain ungapped. Note that the Fermi level in this study has a range as the doping changes its position.}
\end{figure*}

The GdSb$_{x}$Te$_{2-x-\delta}$ series of materials can be viewed as Te-doped relatives of the topological semimetal candidate GdSbTe, the structure of which is shown in Fig.\ \ref{GdSbTe_struct_cdw} \textbf{a} (PbFCl-type structure, $P4/nmm$). This ``parent" material consists of layers of Te and Gd atoms sandwiched between square nets of Sb, Sb-Gd-Te-Te-Gd-Sb; with doping, Te atoms replace Sb atoms in the square net. As the Sb content, $x$, is varied from 1 to 0, the material transforms from an unmodulated tetragonal structure to a modulated orthorhombic superstructure exhibiting multiple CDW wavevectors at $x=0$, GdTe$_{2}$\cite{lei2019charge}. Between $x \sim$ 0.20 \textendash\ 0.80 the material hosts a single CDW wavevector along $b^{*}$, the modulation wavelength of which increases from 4.2 to $\infty$ with decreasing $x$, resulting in an $x$-dependent unidirectional supercell formation along the $b$-axis\cite{lei2019charge}. These CDWs are stable at room temperature and persist for hundreds of K, the CDW transition temperature of GdSb$_{0.46}$Te$_{1.48}$ was found to be approximately 950 K\cite{lei2021band}. The CDW tunability in this region can be related to Fermi surface nesting conditions that change with electron filling (Sb has one fewer electron than Te)\cite{lei2021band}.

The most pronounced consequence of a CDW is the formation of an energy gap across certain regions of the Fermi surface. Gap opening is typically considered to be deleterious in topological semimetals (TSMs) like GdSbTe since it often destroys the non-trivial linear band crossings that define the topological semimetal state. Due to the non-symmorphic symmetry of the crystal structure, though, these materials are known to host symmetry-protected Dirac crossings at certain high-symmetry points in $k$-space, in addition to a number of trivial bands that are also predicted to reside in the vicinity of the Fermi level. Therefore, a CDW forms a gap in the spectrum, but leaves the symmetry-protected Dirac bands untouched. The position of the Fermi level within the symmetry-protected Dirac bands can then be tuned by the Sb content, resulting in a selectively-gapped system which can effectively be gated by the Sb content.\cite{lei2021band}  A schematic band structure outlining this process is shown in Figure 1 \textbf{d}. It should be noted that the location of the Fermi level depends on the doping concentration $x$ of the specific material.

Here we present a survey of the ultrafast dynamics in GdSb$_{x}$Te$_{2-x-\delta}$ ($0.34 \leq\ x \leq\ 0.80$) following photoexcitation of free carriers with mid-infrared (MIR), near-infrared (NIR), and visible light. This range includes material with and without a CDW. NIR and visible light primarily excite above the CDW gap, while MIR light exclusively excites within it. The response in the NIR shows that the CDW compositions and the non-CDW compositions have dynamics with different signs, different timescales, and a different fluence dependence. Spectral features also increase in probe energy with increasing $x$. The responses to MIR and visible excitation were only collected on compounds with the CDW, but these were still able to show the continued evolution of the spectral features into the visible energy range and to differentiate between electronic transitions within and outside of the gap. The fluence dependence of the transient signals shows that CDW compositions near the orthorhombic-tetragonal transition likely undergo a photo-induced phase transition, which is assigned to the transient suppression of the CDW. Ultimately this interpretation is reinforced by the absence of such features in the non-CDW system.

\begin{figure*}
\includegraphics[width=1\textwidth]{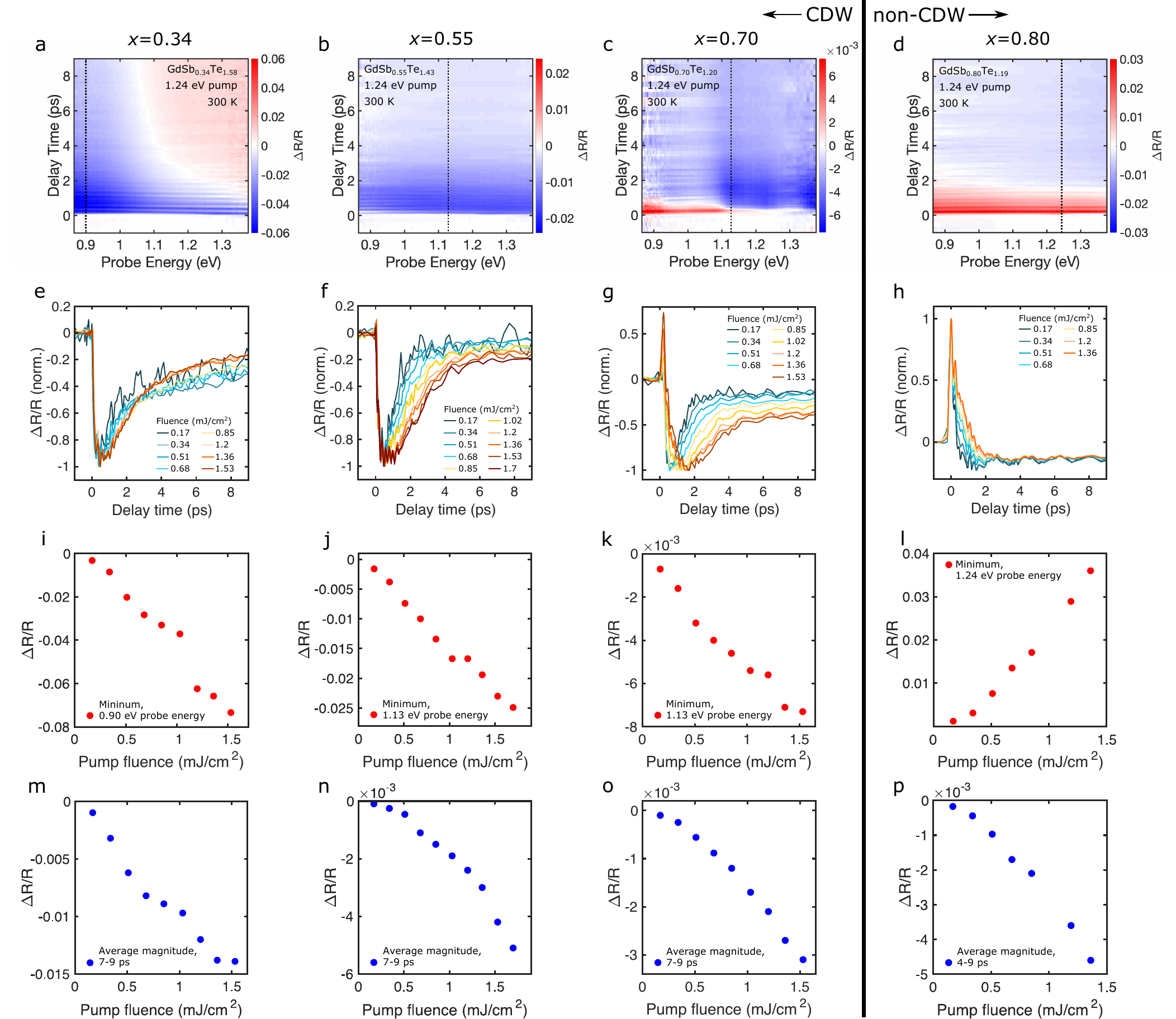}
\caption{\label{GdSbTe_NIR_ultrafast} \textbf{a}-\textbf{d} Transient spectra for GdSb$_{0.34}$Te$_{1.58}$ (\textbf{a}), GdSb$_{0.55}$Te$_{1.43}$ (\textbf{b}), GdSb$_{0.70}$Te$_{1.20}$ (\textbf{c}), and GdSb$_{0.80}$Te$_{1.19}$ (\textbf{d}), tracking the change in the reflected NIR probe pulse after NIR excitation, all at a pump fluence of 1.2 mJ/cm$^{2}$ and at 300 K. \textbf{e}-\textbf{h} Normalized responses at the probe energies denoted by dashed lines in \textbf{a}-\textbf{d}, 0.90 eV (\textbf{e}), 1.13 eV (\textbf{f}), 1.13 eV (\textbf{g}), and 1.24 eV (\textbf{h}), across the range of fluences used. The intensity of the peak (\textbf{i}-\textbf{l}) and the average intensity of the response at later times (\textbf{m}-\textbf{p}; average over 7-9 ps for \textbf{e}-\textbf{g}, 4-9 ps for \textbf{h}) are presented. Additional spectra at other pump fluences can be found in Figures \textbf{S1} \textendash\ \textbf{S4}. }
\end{figure*}

\section{Methods}


The in-depth growth and characterization of these samples is described in Ref. \cite{lei2019charge}, but, in brief: single crystals of GdSb$_{x}$Te$_{2-x-\delta}$ were grown by chemical vapor transport, using iodine as the transport agent. Different starting elemental stoichiometries resulted in various Sb:Te:vacancy ratios; in this work we use samples with $x$ ranging from 0.34 to 0.80. Representative crystals were ground to a powder and analyzed using a STOE STADI P x-ray diffractometer (Mo K${_\alpha1}$ radiation, Ge monochromator). Chemical analysis was done using an XL30 FEG-SEM equipped with an Oxford EDX detector.


NIR pump-NIR probe transient reflectivity experiments were performed with a commercial pump-probe setup (Ultrafast System Helios), in which the 1.55 eV output from a \mbox{1 kHz} regeneratively-amplified Ti:sapphire laser (Coherent Libra) was split to produce both the pump and probe pulses. 1.24 eV narrowband pump pulses were generated in an optical parametric amplifier (OPerA Solo), and broadband probe pulses spanning 0.8 - 1.35 eV were produced by focusing the \mbox{1.55 eV} light in a sapphire crystal.

Visible pump-visible probe and MIR pump-visible probe transient reflectivity experiments were performed on the laser system described in Ref. \cite{montanaro2020visible}. The system is made up of a non-collinear optical parametric amplifier (NOPA, Orpheus-N by Light Conversion) and a twin optical parametric amplifier (TOPAs, Orpheus TWIN by Light Conversion), both pumped by the Pharos Laser (Light Conversion), delivering 400 $\mu$J pulses with 1.2 eV photon energy at 50 kHz. The visible pump (3.1 eV) is obtained by second harmonic generation of the NOPA output in a BBO crystal. The carrier-envelope phase stable MIR pump (0.275 eV) is produced by difference frequency generation in a GaSe crystal of the TOPAs outputs seeded by the same white-light. The broadband visible reflectivity is probed by a white-light supercontinuum (1.4-2.1 eV), generated by focusing the Pharos beam in a 6 mm thick sapphire crystal. The probe pulses are diffracted and acquired by a home-built referenced detection system based on two linear arrays of silicon photodiodes.

All samples were measured in the $ab$ plane. The NIR probe experiments were performed at room temperature, and the samples were housed in an argon-atmosphere chamber because they are air sensitive. The visible probe experiments were performed at 300 K, 150 K, and 10 K, and the samples were held in an air-free closed-cycle helium cryostat.

\section{Results and Discussion}

\subsection{NIR pump and NIR probe}

\begin{figure*}
\includegraphics[width=1\textwidth]{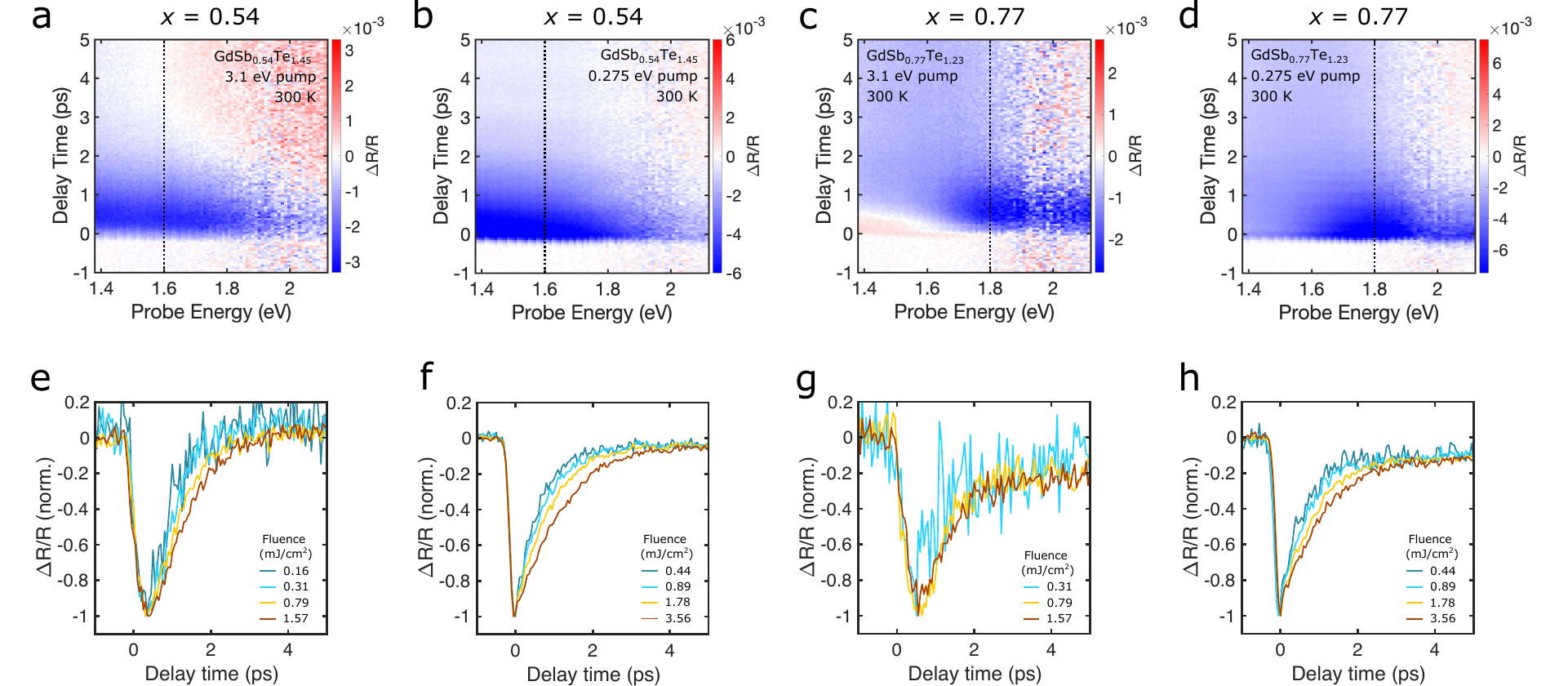}
\caption{\label{GdSbTe_MIR_vis_ultrafast} \textbf{a}-\textbf{d} Transient spectra for GdSb$_{0.54}$Te$_{1.45}$ (\textbf{a}, \textbf{b}) and GdSb$_{0.77}$Te$_{1.23}$ (\textbf{c}, \textbf{d}), tracking the change in the reflected visible probe pulse after excitation by 3.1 eV visible light (\textbf{a} and \textbf{c}, 1.57 mJ/cm$^{2}$) and 0.275 eV MIR light (\textbf{b} and \textbf{d}, 3.56 mJ/cm$^{2}$). All spectra in this figure were collected at 300 K. \textbf{e}-\textbf{h} Normalized responses at the probe energies denoted by dashed lines in \textbf{a}-\textbf{d}, 1.6 eV (\textbf{e}, \textbf{f}) and 1.8 eV (\textbf{g}, \textbf{h}), across the range of fluences used. Additional spectra and dynamics at other pump fluences and at 10 K and 150 K can be found in Figures \textbf{S5} \textendash\ \textbf{S18}. }
\end{figure*}

The room temperature NIR pump-NIR probe spectra for three compounds with a CDW ($x =$ 0.34, 0.55, and 0.70) and one without a CDW ($x =$ 0.80) can be seen in Figure \ref{GdSbTe_NIR_ultrafast} \textbf{a}-\textbf{d}, tracking the transient, or differential, reflectivity ($\Delta R/R$) as a function of probe energy and time delay between the two pulses (additional spectra \textemdash\ at all fluences used \textemdash\ can be found in the Supporting Information, Fig.\ \textbf{S1} \textendash\ \textbf{S4}). Beginning with the compound with the lowest Sb content ($x =$ 0.34, Fig.\ \ref{GdSbTe_NIR_ultrafast} \textbf{a}), and consequently the largest CDW wavevector, the spectrum initially consists of a negative peak centered around 0.90 eV; the closely-related square-net materials ZrSiS and ZrSiSe display a similar initial peak that is due to a transient increase in electronic screening in the excited state\cite{kirby2020}. After 2 ps, the spectrum changes such that the differential signal is negative at low probe energies but positive at higher energies. This final signal is likely the result of lattice heating, and does not reach a nonzero constant value within the delay time window like the subsequent compositions do (Fig.\ \ref{GdSbTe_NIR_ultrafast} \textbf{b}-\textbf{d}, \textbf{f}-\textbf{h}). The transient spectra of the compounds with the next-highest Sb contents, $x =$ 0.55 and 0.70, can be seen in Figure \ref{GdSbTe_NIR_ultrafast} \textbf{b} and \textbf{c}. In contrast to $x =$ 0.34 (Fig.\ \ref{GdSbTe_NIR_ultrafast} \textbf{a}), the spectrum for $x =$ 0.55 is negative over the entire probe range and the main peak is centered around 1.13 eV, slightly higher in energy than in the lower Sb compound. For $x =$ 0.70, the bulk of the NIR response is still negative, (Fig.\ \ref{GdSbTe_NIR_ultrafast} \textbf{c}) but now includes a positive component at low probe energies (0.90 \textendash\ 1.1 eV) and early delay times ($t<1$ ps) that becomes more prominent at high pump fluences. The negative feature found in the $x =$ 0.34 and 0.55 compositions (Fig.\ \ref{GdSbTe_NIR_ultrafast} \textbf{a}, \textbf{b}) appears to shift to still-higher energies for $x =$ 0.70, but there is still a peak at 1.13 eV. In juxtaposition to the lower-Sb orthorhombic CDW phases, the transient response of the ungapped tetragonal $x =$ 0.80 compound is entirely positive in the first 1-2 ps (Fig.\ \ref{GdSbTe_NIR_ultrafast} \textbf{d}). However, it is not clear whether the positive signal is a novel feature of the tetragonal phase or simply a continuation of the feature observed between 0.90 eV and 1.1 eV in $x =$ 0.70; it is conceivable that $x =$ 0.34 and $x =$ 0.55 have similar spectral features at lower probe energies. The scenario in which excitation results in an increase in screening does not account for this change in sign.

The normalized dynamics of the peaks highlighted in each of Figure \ref{GdSbTe_NIR_ultrafast} \textbf{a}-\textbf{d} by a dashed line (0.90 eV (\textbf{a}), 1.13 eV (\textbf{b}), 1.13 eV (\textbf{c}), and 1.24 eV (\textbf{d}) are shown in the main panels of Figures \ref{GdSbTe_NIR_ultrafast} \textbf{e}-\textbf{h} for a range of pump fluences. For the compound with the largest CDW wavevector, $x =$ 0.34, the time constant for the recovery, hereafter referred to as the ``lifetime", increases as the fluence is increased from 0.17 mJ/cm$^{2}$ to 1.53 mJ/cm$^{2}$ (Fig.\ \ref{GdSbTe_NIR_ultrafast} \textbf{e}). Coherent phonons are also excited at 125 cm$^{-1}$ and 160 cm$^{-1}$, and are due to the presence of Te-O impurities on the sample surface\cite{lei2020high, wang2022axial} which occur even when the samples are housed in an Ar glove box, measured in an Ar atmosphere, or even measured under vacuum in a cryostat; they are ubiquitous but extraneous to the bulk electronic dynamics. For $x =$ 0.55 (Fig.\ \ref{GdSbTe_NIR_ultrafast} \textbf{f}), the lifetime of the initial photoexcited state increases with increasing fluence, similar to $x =$ 0.34 \textemdash\ but unlike the lower-Sb composition achieves a metastable constant signal within the delay time window, in which the temperature of the lattice is likely elevated out of equilibrium. Similar to $x =$ 0.34 and $x =$ 0.55, the lifetime of the 1.13 eV peak in $x =$ 0.70 increases with increasing fluence (Fig.\ \ref{GdSbTe_NIR_ultrafast} \textbf{g}), and, like $x =$ 0.55, the response attains a constant quasi-equilibrium state within a few ps. The position of this peak in time (the ``rise time'') also increases with increasing fluence, exhibiting threshold-like behavior (Fig.\ \ref{GdSbTe_NIR_ultrafast} \textbf{g}). This is in contrast to the lower Sb samples, in which the peak broadens slightly with increasing fluence but the position of it in time does not significantly change (Fig.\ \ref{GdSbTe_NIR_ultrafast} \textbf{e}, \textbf{f}). This is discussed further in the \textbf{Fluence and temperature dependence} part of this section. In the non-CDW compound, $x =$ 0.80, this initial decay happens much more quickly (Fig.\ \ref{GdSbTe_NIR_ultrafast} \textbf{h}) than in the partially gapped phases with lower Sb contents (Fig.\ \ref{GdSbTe_NIR_ultrafast} \textbf{e}-\textbf{g}), which is consistent with this composition being ungapped, but it still increases with increasing fluence. In addition to the endemic Te-O phonon modes, this composition displays very clear low-frequency coherent phonons (17 cm$^{-1}$) that do not dephase within the delay time window.

In addition to transient signals from the gapped and ungapped compounds having opposite signs, there are also differences in the fluence dependence that depend on the presence or absence of the CDW (Fig.\ \ref{GdSbTe_NIR_ultrafast} \textbf{i}-\textbf{p}). In all of the compounds in the CDW phase ($x =$ 0.34, 0.55, and 0.70), the intensity of the negative peak (Fig.\ \ref{GdSbTe_NIR_ultrafast} \textbf{i}-\textbf{k}, probe energies of 0.90 eV, 1.13 eV, and 1.13 eV, respectively) increases linearly with fluence, whereas in the ungapped phase ($x =$ 0.80, Fig.\ \ref{GdSbTe_NIR_ultrafast} \textbf{l}) the magnitude increases superlinearly with fluence. The quasi-equilibrium response at later times does not follow this trend (Fig.\ \ref{GdSbTe_NIR_ultrafast} \textbf{m}-\textbf{p}); in $x =$ 0.80, the constant signal increases superlinearly with increasing fluence, like the initial peak. The plateaus in $x =$ 0.55 and 0.70 also, however, increase superlinearly in magnitude with fluence, while the $x =$ 0.34 composition does not reach such a plateau within our delay time window. If a superlinear response is characteristic of the tetragonal phase, this could suggest that, after the initial excitation, the $x =$ 0.55 and 0.70 compounds find themselves in a metastable state that responds more similarly to the tetragonal phase, in which the CDW is at least partially suppressed. The q vectors decrease in magnitude with increasing Sb content, so it appears reasonable that the orthorhombic compounds with higher Sb contents will have a lower energy barrier to CDW suppression than lower Sb contents, like $x =$ 0.34.

\begin{figure}
\includegraphics[width=0.5\textwidth]{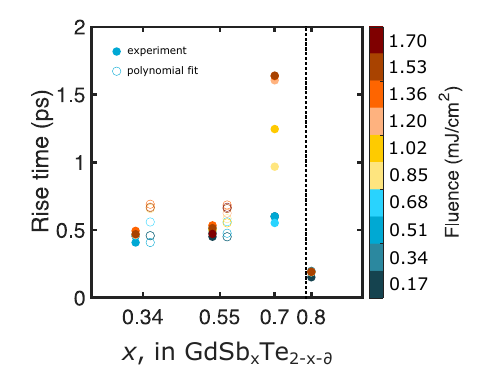}
\caption{\label{GdSbTe_risetime} Evolution of the rise time in the compositions studied in the NIR. In $x =$ 0.70, there is a threshold fluence between 0.60 mJ/cm$^{2}$ and 0.85 mJ/cm$^{2}$, above which the delay minimum increases linearly with fluence, before plateauing around 1.2 mJ/cm$^{2}$; although the negative peaks broaden slightly in the other compositions, they do not display such drastic fluence dependence. The transition from the orthorhombic phase to the tetragonal phase is denoted with a dashed line. In $x =$ 0.34 and 0.55, the minima of the transients and the minima from high-order polynomial fits are included, to minimize potential artefacts from the Te-O impurity coherent phonon response. }
\end{figure}

\subsection{Visible pump and visible probe, and MIR pump and visible probe}

Since the spectral features observed with the NIR probe pulse shift with Sb content to higher energy (Fig.\ \ref{GdSbTe_NIR_ultrafast} \textbf{a}-\textbf{d}), the continuation of this trend into the visible region was investigated with further transient reflectivity measurements using a visible probe pulse. Compounds with $x =$ 0.54 and 0.77 (both CDW-hosting) were thus studied in this region, with a 3.1 eV pump pulse (Fig.\ \ref{GdSbTe_MIR_vis_ultrafast} \textbf{a}, \textbf{c}), as well as with a 0.275 eV MIR pump pulse (Fig.\ \ref{GdSbTe_MIR_vis_ultrafast} \textbf{b}, \textbf{d}). With $x =$ 0.54, the broad negative peak observed in $x =$ 0.55 (Fig.\ \ref{GdSbTe_NIR_ultrafast} \textbf{a}) extends into the visible probe window (Fig.\ \ref{GdSbTe_MIR_vis_ultrafast} \textbf{a}, \textbf{b} \textbf{e}, \textbf{f}, \textbf{S5} \textendash\ \textbf{S11}), displaying a minimum around 1.6 eV. This broad peak occurs regardless of whether the material is pumped with 0.275 eV or 3.1 eV light (Fig.\ \ref{GdSbTe_MIR_vis_ultrafast} \textbf{a}, \textbf{b}), and is likely the result of lattice heating. Additionally, when pumped at 0.275 eV, this composition displays low-frequency coherent phonons similar to those observed in $x =$ 0.80, although at a slightly higher frequency, 33 cm$^{-1}$.

In $x =$ 0.77, there are larger differences between the response to the 0.275 eV and the 3.1 eV pumps at early delay times (Fig.\ \ref{GdSbTe_MIR_vis_ultrafast} \textbf{c}, \textbf{d}, \textbf{g}, \textbf{h}, \textbf{S12} \textendash\ \textbf{S18}). The main peak is centered around 1.8 eV in both situations, but 3.1 eV excitation leads to a positive feature within the first ps like that seen in $x =$ 0.70 (Fig.\ \ref{GdSbTe_NIR_ultrafast} \textbf{d}), but at higher energy (1.4-1.6 eV). This follows the pattern of spectral features increasing in energy with increasing Sb content. This feature is absent in the response when pumped at 0.275 eV and 0.077 eV (Fig.\ \textbf{S19}), and as such is likely a consequence of above-gap excitation.

For the CDW compound $x =$ 0.77, a spectral feature associated with transitions well above the CDW gap has been identified. That this feature is also present for $x =$ 0.70 following 1.24 eV excitation provides some level of upper and lower bounds for the energy scale of the CDW: 0.276 eV $<$ $2\Delta_{CDW}$ $<$ 1.24 eV. Although the 3.1 eV pump is roughly 11x the energy of the 0.276 eV pump and has been shown to produce different initial excited states, after 2 ps each compound ends up in the same quasi-equilibrium `hot lattice' state.

\begin{figure}
\includegraphics[width=0.8\textwidth]{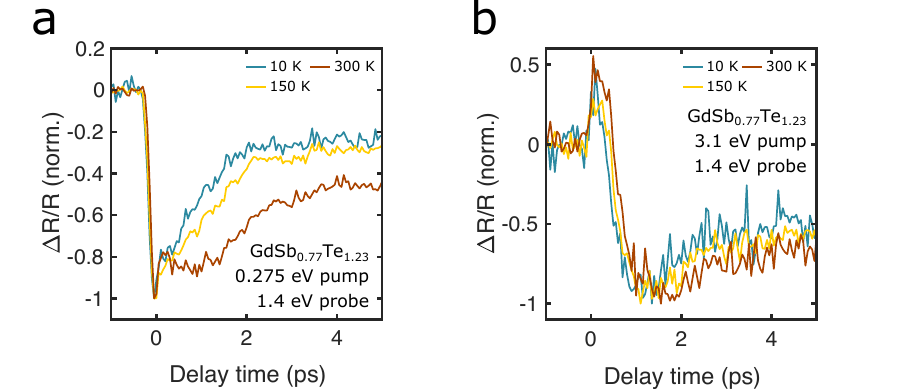}
\caption{\label{GdSbTe_temp} \textbf{a} In $x =$ 0.77, following 0.275 eV excitation, a second minimum grows in with increasing sample temperature, and at each temperature with increasing pump fluence. \textbf{b} $x =$ 0.77 also displays a moving minimum when pumped at 3.1 eV. Full spectra from \textbf{a} and \textbf{b} can be found in Figures \textbf{S12} \textendash\ \textbf{S17}. }
\end{figure}

\subsection{Fluence and temperature dependence}

As previously mentioned, the response in all of these compositions has a strong pump fluence dependence. In every composition, the lifetime increases with increasing excitation density. Although gapped systems are known to have increasing lifetimes as the gap size decreases due to an increased phonon reabsorption rate (often described through the lens of the Rothwarf-Taylor model for superconductors), this effect persists above the CDW transition composition (Fig.\ \ref{GdSbTe_NIR_ultrafast} \textbf{h}) so is likely due to excitation density effects. However, there are fluence-dependent effects observed solely in the CDW compounds approaching the transition composition, $x =$ 0.70 and 0.77; these materials also have the smallest q vectors and are thought to have the lowest CDW melting temperatures out the series studied here. In $x =$ 0.70, this fluence dependence is manifested through the delay time at which the 1.13 eV peak attains its minimum $\Delta R/R$ signal, the `rise time' (Fig.\ \ref{GdSbTe_NIR_ultrafast} \textbf{g} and \ref{GdSbTe_risetime}). Below 0.85 mJ/cm$^{2}$, the peak occurs around 0.60 ps, independent of fluence. As the fluence increases above 0.85 mJ/cm$^{2}$, it takes the response longer and longer to reach its minimum, with the peak progressing roughly linearly in time with increasing fluence until 1.2 mJ/cm$^{2}$, at which point it appears to saturate, over 1 ps later than at low fluences.

This threshold-like behavior in the time it takes the system to become maximally out of equilibrium, which in CDW systems can be thought of as the time it takes to suppress the modulation, is often indicative of a photo-induced phase transition. A similar change in the rise time of the response was observed in LaTe$_{3}$ (``dynamical slowing down", in which the lifetime of the CDW order parameter increases around a critical fluence) and was attributed to the transient optical suppression of the CDW\cite{zong2019dynamical}. Although partial suppression of the CDW in $x =$ 0.70 is possible, it is unlikely that we are observing complete CDW melting (with LaTe$_{3}$, the authors were able to clearly traverse the increase and subsequent decrease in rise time with their fluence range), and the exact nature of this transition remains unknown. The other compositions display slight increases in their rise times with increasing fluence but no threshold-like behavior (Fig.\ \ref{GdSbTe_risetime}). Despite the fact that a Rothwarf-Taylor picture seems to be an inadequate descriptor of the transient dynamics of each individual composition \textemdash\ because the lifetimes increase with fluence on both sides of the CDW transition composition \textemdash\ the spread of the rise times with Sb content (Fig.\ \ref{GdSbTe_risetime}) follows the typical path of an order parameter around T$_{C}$. As the $x$ increases from 0.34 to 0.70, the rise times and the spread in the rise times increase; far below the transition (e.g., $x =$ 0.34 and 0.55) the changes are modest, yet near the transition the rise times increase dramatically. On the other side of the transition composition, i.e. $x =$ 0.80, the rise times are appreciably lower than below the transition.

The MIR and visible pump experiments on $x =$ 0.54 and 0.77 were performed at temperatures down to 10 K, which is below the magnetic ordering temperature in these materials\cite{lei2019charge}. The Sb content-dependent T$_{C}$ for the CDW in these materials is far above room temperature, so performing measurements at cryogenic temperatures will not provide information towards the dynamics of a particular composition above and below its CDW T$_{C}$, but it can give indications that the excess energy deposited in the system is contributing to a photo-induced phase transition. When $x =$ 0.77 is pumped at 0.275 eV, the response at 1.4 eV presents a secondary minimum that becomes more prominent with increasing temperature and with increasing fluence. This second minimum also occurs at later delay times with increasing temperature and fluence, a similar effect to that observed in $x =$ 0.70 at 300 K with fluence. When pumped at 3.1 eV, the transients at 1.4 eV also show some fluence dependence, in which the rise time stretches from $\approx$ 1 ps to 2 ps as the temperature is raised from 10 K to 300 K (Fig.\ \ref{GdSbTe_temp} \textbf{b}). That these photo-induced phenomena become more prominent with increasing fluence and especially increasing temperature suggests that there is some barrier to physical change that is easier for the material to surmount when it has more excess energy. That the compositions that exhibit strong fluence-dependent dynamics, $x =$ 0.70 and 0.77, are also those nearest the transition composition indicate that the photo-induced effects are likely related to the nearby tetragonal unmodulated phase.

\section{Conclusions}

To conclude, we have presented the first systematic ultrafast spectroscopic study of a material tuned between undistorted and CDW phases. Specifically we studied the GdSb$_{x}$Te$_{2-x-\delta}$ series of materials, focusing on their optical response in the NIR and visible, while pumping with MIR, NIR or visible light. As the Sb content increases, transient spectral features shift to higher probe energies, showing that the optoelectronic properties of the material similarly evolve continuously with Sb content. In compounds with the CDW, the initial response is negative, with lifetimes that increase with pump fluence. In the compound without a CDW, the transient response changes sign and displays much faster relaxation. This clearly demonstrates the long rise and recovery lifetimes result from the CDW destruction and reformation. Furthermore the ultrafast spectral response measured in the NIR is similar for excitations both intra-gap and well above the CDW gap. This is consistent with either the CDW coupling to the overall bandstructure, and/or ultrafast response primarily resulting from sample heating. In the CDW compounds close to the transition composition ($x =$ 0.70 and 0.77), there are temperature- and fluence-dependent phenomena that might be indicative of some photo-induced change to the electronic structure of the material \textemdash\ such as the suppression of the CDW \textemdash\ that warrants additional investigation. We hope that the work presented here can lay the foundation from which to study the photo-induced phenomena in this complex tunable correlated system further.

\begin{acknowledgement}

This work was supported by AFSOR, grant FA9550-20-1-0246. Additional support came from NSF through the Princeton Center for Complex Materials, a Materials Research Science and Engineering Center DMR-2011750, by Princeton University through the Princeton Catalysis Initiative, and by the Gordon and Betty Moore Foundation through Grant GBMF9064 to L.M.S. G.D.S. is a CIFAR Fellow in the Bio-Inspired Energy Program. D.F. was supported by the European Commission through the European Research Council (ERC), Project INCEPT, Grant 677488. Work by I.P. and P.N. is supported by the Quantum Science Center (QSC), a National Quantum Information Science Research Center of the U.S. Department of Energy (DOE). P.N. acknowledges support from a CIFAR BSE Catalyst Program that has facilitated interactions and collaborations with G.D.S. The authors also acknowledge the use of Princeton's Imaging and Analysis Center, which is partially supported by the Princeton Center for Complex Materials, a National Science Foundation (NSF)-MRSEC program (DMR-2011750). 

\end{acknowledgement}

\begin{suppinfo}

Additional NIR pump-NIR probe spectra at 300 K, MIR pump-visible probe and visible pump-visible probe spectra at 10, 150, and 300 K, and MIR pump-visible probe spectra for the 0.077 eV MIR pump

\end{suppinfo}


\begin{mcitethebibliography}{42}
\providecommand*\natexlab[1]{#1}
\providecommand*\mciteSetBstSublistMode[1]{}
\providecommand*\mciteSetBstMaxWidthForm[2]{}
\providecommand*\mciteBstWouldAddEndPuncttrue
  {\def\EndOfBibitem{\unskip.}}
\providecommand*\mciteBstWouldAddEndPunctfalse
  {\let\EndOfBibitem\relax}
\providecommand*\mciteSetBstMidEndSepPunct[3]{}
\providecommand*\mciteSetBstSublistLabelBeginEnd[3]{}
\providecommand*\EndOfBibitem{}
\mciteSetBstSublistMode{f}
\mciteSetBstMaxWidthForm{subitem}{(\alph{mcitesubitemcount})}
\mciteSetBstSublistLabelBeginEnd
  {\mcitemaxwidthsubitemform\space}
  {\relax}
  {\relax}

\bibitem[Gr{\"u}ner(1988)]{gruner1988dynamics}
Gr{\"u}ner,~G. {The dynamics of charge-density waves}. \emph{Rev.\ Mod.\ Phys.}
  \textbf{1988}, \emph{60}, 1129\relax
\mciteBstWouldAddEndPuncttrue
\mciteSetBstMidEndSepPunct{\mcitedefaultmidpunct}
{\mcitedefaultendpunct}{\mcitedefaultseppunct}\relax
\EndOfBibitem
\bibitem[Demsar \latin{et~al.}(1999)Demsar, Biljakovi{\'c}, and
  Mihailovic]{demsar1999single}
Demsar,~J.; Biljakovi{\'c},~K.; Mihailovic,~D. {Single particle and collective
  excitations in the one-dimensional charge density wave solid
  K$_{0.3}$MoO$_{3}$ probed in real time by femtosecond spectroscopy}.
  \emph{Phys.\ Rev.\ Lett.} \textbf{1999}, \emph{83}, 800\relax
\mciteBstWouldAddEndPuncttrue
\mciteSetBstMidEndSepPunct{\mcitedefaultmidpunct}
{\mcitedefaultendpunct}{\mcitedefaultseppunct}\relax
\EndOfBibitem
\bibitem[Demsar \latin{et~al.}(2002)Demsar, Forr{\'o}, Berger, and
  Mihailovic]{demsar2002femtosecond}
Demsar,~J.; Forr{\'o},~L.; Berger,~H.; Mihailovic,~D. {Femtosecond snapshots of
  gap-forming charge-density-wave correlations in quasi-two-dimensional
  dichalcogenides 1T-TaS$_{2}$ and 2H-TaSe$_{2}$}. \emph{Phys.\ Rev.\ B}
  \textbf{2002}, \emph{66}, 041101\relax
\mciteBstWouldAddEndPuncttrue
\mciteSetBstMidEndSepPunct{\mcitedefaultmidpunct}
{\mcitedefaultendpunct}{\mcitedefaultseppunct}\relax
\EndOfBibitem
\bibitem[Tomeljak \latin{et~al.}(2009)Tomeljak, Schaefer, St{\"a}dter, Beyer,
  Biljakovic, and Demsar]{tomeljak2009dynamics}
Tomeljak,~A.; Schaefer,~H.; St{\"a}dter,~D.; Beyer,~M.; Biljakovic,~K.;
  Demsar,~J. {Dynamics of photoinduced charge-density-wave to metal phase
  transition in K$_{0.3}$MoO$_{3}$}. \emph{Phys.\ Rev.\ Lett.} \textbf{2009},
  \emph{102}, 066404\relax
\mciteBstWouldAddEndPuncttrue
\mciteSetBstMidEndSepPunct{\mcitedefaultmidpunct}
{\mcitedefaultendpunct}{\mcitedefaultseppunct}\relax
\EndOfBibitem
\bibitem[Hellmann \latin{et~al.}(2010)Hellmann, Beye, Sohrt, Rohwer,
  Sorgenfrei, Redlin, Kall{\"a}ne, Marczynski-B{\"u}hlow, Hennies, Bauer,
  F{\"o}hlisch, Kipp, Wurth, and Rossnagel]{hellmann2010ultrafast}
Hellmann,~S.; Beye,~M.; Sohrt,~C.; Rohwer,~T.; Sorgenfrei,~F.; Redlin,~H.;
  Kall{\"a}ne,~M.; Marczynski-B{\"u}hlow,~M.; Hennies,~F.; Bauer,~M.
  \latin{et~al.}  {Ultrafast Melting of a Charge-Density Wave in the Mott
  Insulator 1T-TaS$_{2}$}. \emph{Phys.\ Rev.\ Lett.} \textbf{2010}, \emph{105},
  187401\relax
\mciteBstWouldAddEndPuncttrue
\mciteSetBstMidEndSepPunct{\mcitedefaultmidpunct}
{\mcitedefaultendpunct}{\mcitedefaultseppunct}\relax
\EndOfBibitem
\bibitem[M{\"o}hr-Vorobeva \latin{et~al.}(2011)M{\"o}hr-Vorobeva, Johnson,
  Beaud, Staub, De~Souza, Milne, Ingold, Demsar, Sch{\"a}fer, and
  Titov]{mohr2011nonthermal}
M{\"o}hr-Vorobeva,~E.; Johnson,~S.~L.; Beaud,~P.; Staub,~U.; De~Souza,~R.;
  Milne,~C.; Ingold,~G.; Demsar,~J.; Sch{\"a}fer,~H.; Titov,~A. {Nonthermal
  melting of a charge density wave in TiSe$_{2}$}. \emph{Phys.\ Rev.\ Lett.}
  \textbf{2011}, \emph{107}, 036403\relax
\mciteBstWouldAddEndPuncttrue
\mciteSetBstMidEndSepPunct{\mcitedefaultmidpunct}
{\mcitedefaultendpunct}{\mcitedefaultseppunct}\relax
\EndOfBibitem
\bibitem[Lin \latin{et~al.}(2020)Lin, Shi, Wang, Zhang, Liu, Hu, Dong, Wu, and
  Wang]{lin2020optical}
Lin,~T.; Shi,~L.~Y.; Wang,~Z.~X.; Zhang,~S.~J.; Liu,~Q.~M.; Hu,~T.~C.;
  Dong,~T.; Wu,~D.; Wang,~N.~L. {Optical spectroscopy and ultrafast pump-probe
  study on Bi$_{2}$Rh$_{3}$Se$_{2}$: Evidence for charge density wave order
  formation}. \emph{Phys.\ Rev.\ B} \textbf{2020}, \emph{101}, 205112\relax
\mciteBstWouldAddEndPuncttrue
\mciteSetBstMidEndSepPunct{\mcitedefaultmidpunct}
{\mcitedefaultendpunct}{\mcitedefaultseppunct}\relax
\EndOfBibitem
\bibitem[Anikin \latin{et~al.}(2020)Anikin, Schaller, Wiederrecht, Margine,
  Mazin, and Karapetrov]{anikin2020ultrafast}
Anikin,~A.; Schaller,~R.~D.; Wiederrecht,~G.~P.; Margine,~E.~R.; Mazin,~I.~I.;
  Karapetrov,~G. {Ultrafast dynamics in the high-symmetry and in the charge
  density wave phase of 2H-NbSe$_{2}$}. \emph{Phys.\ Rev.\ B} \textbf{2020},
  \emph{102}, 205139\relax
\mciteBstWouldAddEndPuncttrue
\mciteSetBstMidEndSepPunct{\mcitedefaultmidpunct}
{\mcitedefaultendpunct}{\mcitedefaultseppunct}\relax
\EndOfBibitem
\bibitem[Monney \latin{et~al.}(2016)Monney, Puppin, Nicholson, Hoesch, Chapman,
  Springate, Berger, Magrez, Cacho, Ernstorfer, and Wolf]{monney2016revealing}
Monney,~C.; Puppin,~M.; Nicholson,~C.~W.; Hoesch,~M.; Chapman,~R.~T.;
  Springate,~E.; Berger,~H.; Magrez,~A.; Cacho,~C.; Ernstorfer,~R.
  \latin{et~al.}  {Revealing the role of electrons and phonons in the ultrafast
  recovery of charge density wave correlations in 1T-TiSe$_{2}$}. \emph{Phys.\
  Rev.\ B} \textbf{2016}, \emph{94}, 165165\relax
\mciteBstWouldAddEndPuncttrue
\mciteSetBstMidEndSepPunct{\mcitedefaultmidpunct}
{\mcitedefaultendpunct}{\mcitedefaultseppunct}\relax
\EndOfBibitem
\bibitem[Stojchevska \latin{et~al.}(2014)Stojchevska, Vaskivskyi, Mertelj,
  Kusar, Svetin, Brazovskii, and Mihailovic]{stojchevska2014ultrafast}
Stojchevska,~L.; Vaskivskyi,~I.; Mertelj,~T.; Kusar,~P.; Svetin,~D.;
  Brazovskii,~S.; Mihailovic,~D. {Ultrafast switching to a stable hidden
  quantum state in an electronic crystal}. \emph{Science} \textbf{2014},
  \emph{344}, 177--180\relax
\mciteBstWouldAddEndPuncttrue
\mciteSetBstMidEndSepPunct{\mcitedefaultmidpunct}
{\mcitedefaultendpunct}{\mcitedefaultseppunct}\relax
\EndOfBibitem
\bibitem[Sun \latin{et~al.}(2018)Sun, Sun, Zhu, Tian, Yang, and
  Li]{sun2018hidden}
Sun,~K.; Sun,~S.; Zhu,~C.; Tian,~H.; Yang,~H.; Li,~J. {Hidden CDW states and
  insulator-to-metal transition after a pulsed femtosecond laser excitation in
  layered chalcogenide 1T-TaS$_{2-x}$Se$_{x}$}. \emph{Sci.\ Adv.}
  \textbf{2018}, \emph{4}, eaas9660\relax
\mciteBstWouldAddEndPuncttrue
\mciteSetBstMidEndSepPunct{\mcitedefaultmidpunct}
{\mcitedefaultendpunct}{\mcitedefaultseppunct}\relax
\EndOfBibitem
\bibitem[Yoshikawa \latin{et~al.}(2021)Yoshikawa, Suganuma, Matsuoka, Tanaka,
  Hemme, Cazayous, Gallais, Nakano, Iwasa, and Shimano]{yoshikawa2021ultrafast}
Yoshikawa,~N.; Suganuma,~H.; Matsuoka,~H.; Tanaka,~Y.; Hemme,~P.; Cazayous,~M.;
  Gallais,~Y.; Nakano,~M.; Iwasa,~Y.; Shimano,~R. {Ultrafast switching to an
  insulating-like metastable state by amplitudon excitation of a charge density
  wave}. \emph{Nat.\ Phys.} \textbf{2021}, \emph{17}, 909--914\relax
\mciteBstWouldAddEndPuncttrue
\mciteSetBstMidEndSepPunct{\mcitedefaultmidpunct}
{\mcitedefaultendpunct}{\mcitedefaultseppunct}\relax
\EndOfBibitem
\bibitem[Maklar \latin{et~al.}(2022)Maklar, Dong, Sarkar, Gerasimenko,
  Pincelli, Beaulieu, Kirchmann, Sobota, Yang, Leuenberger, Moore, Shen, Wolf,
  Mihailovic, Ernstorfer, and Rettig]{maklar2022coherent}
Maklar,~J.; Dong,~S.; Sarkar,~J.; Gerasimenko,~Y.~A.; Pincelli,~T.;
  Beaulieu,~S.; Kirchmann,~P.~S.; Sobota,~J.~A.; Yang,~S.-L.; Leuenberger,~D.
  \latin{et~al.}  {Coherent Light Control of a Metastable Hidden Phase}.
  \emph{arXiv preprint arXiv:2206.03788} \textbf{2022}, \relax
\mciteBstWouldAddEndPunctfalse
\mciteSetBstMidEndSepPunct{\mcitedefaultmidpunct}
{}{\mcitedefaultseppunct}\relax
\EndOfBibitem
\bibitem[Sagar \latin{et~al.}(2007)Sagar, Tsvetkov, Fausti, van Smaalen, and
  van Loosdrecht]{sagar2007coherent}
Sagar,~D.~M.; Tsvetkov,~A.~A.; Fausti,~D.; van Smaalen,~S.; van Loosdrecht,~P.
  H.~M. Coherent amplitudon generation in blue bronze through ultrafast
  interband quasi-particle decay. \emph{J.\ Phys.: Condens.\ Matter}
  \textbf{2007}, \emph{19}, 346208\relax
\mciteBstWouldAddEndPuncttrue
\mciteSetBstMidEndSepPunct{\mcitedefaultmidpunct}
{\mcitedefaultendpunct}{\mcitedefaultseppunct}\relax
\EndOfBibitem
\bibitem[Sagar \latin{et~al.}(2008)Sagar, Fausti, Yue, Kuntscher, van Smaalen,
  and van Loosdrecht]{sagar2008raman}
Sagar,~D.~M.; Fausti,~D.; Yue,~S.; Kuntscher,~C.~A.; van Smaalen,~S.; van
  Loosdrecht,~P. H.~M. {A Raman study of the charge-density-wave state in
  A$_{0.3}$MoO$_{3}$ (A= K, Rb)}. \emph{New J.\ Phys.} \textbf{2008},
  \emph{10}, 023043\relax
\mciteBstWouldAddEndPuncttrue
\mciteSetBstMidEndSepPunct{\mcitedefaultmidpunct}
{\mcitedefaultendpunct}{\mcitedefaultseppunct}\relax
\EndOfBibitem
\bibitem[Ren \latin{et~al.}(2004)Ren, Xu, and L{\"u}pke]{ren2004ultrafast}
Ren,~Y.; Xu,~Z.; L{\"u}pke,~G. {Ultrafast collective dynamics in the
  charge-density-wave conductor K$_{0.3}$MoO$_{3}$}. \emph{Chem.\ Phys.}
  \textbf{2004}, \emph{120}, 4755--4758\relax
\mciteBstWouldAddEndPuncttrue
\mciteSetBstMidEndSepPunct{\mcitedefaultmidpunct}
{\mcitedefaultendpunct}{\mcitedefaultseppunct}\relax
\EndOfBibitem
\bibitem[Sacchetti \latin{et~al.}(2006)Sacchetti, Degiorgi, Giamarchi, Ru, and
  Fisher]{sacchetti2006chemical}
Sacchetti,~A.; Degiorgi,~L.; Giamarchi,~T.; Ru,~N.; Fisher,~I.~R. {Chemical
  pressure and hidden one-dimensional behavior in rare-earth tri-telluride
  charge-density wave compounds}. \emph{Phys.\ Rev.\ B} \textbf{2006},
  \emph{74}, 125115\relax
\mciteBstWouldAddEndPuncttrue
\mciteSetBstMidEndSepPunct{\mcitedefaultmidpunct}
{\mcitedefaultendpunct}{\mcitedefaultseppunct}\relax
\EndOfBibitem
\bibitem[Yusupov \latin{et~al.}(2008)Yusupov, Mertelj, Chu, Fisher, and
  Mihailovic]{yusupov2008single}
Yusupov,~R.~V.; Mertelj,~T.; Chu,~J.-H.; Fisher,~I.~R.; Mihailovic,~D.
  {Single-particle and collective mode couplings associated with 1- and
  2-directional electronic ordering in metallic RTe$_{3}$ (R= Ho, Dy, Tb)}.
  \emph{Phys.\ Rev.\ Lett.} \textbf{2008}, \emph{101}, 246402\relax
\mciteBstWouldAddEndPuncttrue
\mciteSetBstMidEndSepPunct{\mcitedefaultmidpunct}
{\mcitedefaultendpunct}{\mcitedefaultseppunct}\relax
\EndOfBibitem
\bibitem[Schmitt \latin{et~al.}(2008)Schmitt, Kirchmann, Bovensiepen, Moore,
  Rettig, Krenz, Chu, Ru, Perfetti, Lu, Wolf, Fisher, and
  Shen]{schmitt2008transient}
Schmitt,~F.; Kirchmann,~P.~S.; Bovensiepen,~U.; Moore,~R.~G.; Rettig,~L.;
  Krenz,~M.; Chu,~J.-H.; Ru,~N.; Perfetti,~L.; Lu,~D.~H. \latin{et~al.}
  {Transient electronic structure and melting of a charge density wave in
  TbTe$_3$}. \emph{Science} \textbf{2008}, \emph{321}, 1649--1652\relax
\mciteBstWouldAddEndPuncttrue
\mciteSetBstMidEndSepPunct{\mcitedefaultmidpunct}
{\mcitedefaultendpunct}{\mcitedefaultseppunct}\relax
\EndOfBibitem
\bibitem[Hu \latin{et~al.}(2011)Hu, Zheng, Yuan, Dong, Cheng, Chen, and
  Wang]{hu2011}
Hu,~B.~F.; Zheng,~P.; Yuan,~R.~H.; Dong,~T.; Cheng,~B.; Chen,~Z.~G.;
  Wang,~N.~L. {Optical spectroscopy study on CeTe$_{3}$: Evidence for multiple
  charge-density-wave orders}. \emph{Phys.\ Rev.\ B} \textbf{2011}, \emph{83},
  155113\relax
\mciteBstWouldAddEndPuncttrue
\mciteSetBstMidEndSepPunct{\mcitedefaultmidpunct}
{\mcitedefaultendpunct}{\mcitedefaultseppunct}\relax
\EndOfBibitem
\bibitem[Hu \latin{et~al.}(2011)Hu, Cheng, Yuan, Dong, Fang, Guo, Chen, Zheng,
  Shi, and Wang]{hu2011optical}
Hu,~B.~F.; Cheng,~B.; Yuan,~R.~H.; Dong,~T.; Fang,~A.~F.; Guo,~W.~T.;
  Chen,~Z.~G.; Zheng,~P.; Shi,~Y.~G.; Wang,~N.~L. {Optical study of the
  multiple charge-density-wave transitions in ErTe$_{3}$}. \emph{Phys.\ Rev.\
  B} \textbf{2011}, \emph{84}, 155132\relax
\mciteBstWouldAddEndPuncttrue
\mciteSetBstMidEndSepPunct{\mcitedefaultmidpunct}
{\mcitedefaultendpunct}{\mcitedefaultseppunct}\relax
\EndOfBibitem
\bibitem[Lavagnini \latin{et~al.}(2012)Lavagnini, Pfuner, Monnier, Degiorgi,
  Eiter, Tassini, Muschler, Hackl, Chu, Ru, Shin, and
  Fisher]{lavagnini2012infrared}
Lavagnini,~M.; Pfuner,~F.; Monnier,~R.; Degiorgi,~L.; Eiter,~H.-M.;
  Tassini,~L.; Muschler,~B.; Hackl,~R.; Chu,~J.-H.; Ru,~N. \latin{et~al.}
  {Infrared and Raman investigation of the charge-density wave state in
  rare-earth tri-telluride compounds}. \emph{Phys.\ B: Condens.\ Matter}
  \textbf{2012}, \emph{407}, 1864--1867\relax
\mciteBstWouldAddEndPuncttrue
\mciteSetBstMidEndSepPunct{\mcitedefaultmidpunct}
{\mcitedefaultendpunct}{\mcitedefaultseppunct}\relax
\EndOfBibitem
\bibitem[Hu \latin{et~al.}(2014)Hu, Cheng, Yuan, Dong, and
  Wang]{hu2014coexistence}
Hu,~B.~F.; Cheng,~B.; Yuan,~R.~H.; Dong,~T.; Wang,~N.~L. {Coexistence and
  competition of multiple charge-density-wave orders in rare-earth
  tritellurides}. \emph{Phys.\ Rev.\ B} \textbf{2014}, \emph{90}, 085105\relax
\mciteBstWouldAddEndPuncttrue
\mciteSetBstMidEndSepPunct{\mcitedefaultmidpunct}
{\mcitedefaultendpunct}{\mcitedefaultseppunct}\relax
\EndOfBibitem
\bibitem[Chen \latin{et~al.}(2014)Chen, Hu, Dong, and Wang]{chen2014revealing}
Chen,~R.~Y.; Hu,~B.~F.; Dong,~T.; Wang,~N.~L. {Revealing multiple
  charge-density-wave orders in TbTe$_{3}$ by optical conductivity and
  ultrafast pump-probe experiments}. \emph{Phys.\ Rev.\ B} \textbf{2014},
  \emph{89}, 075114\relax
\mciteBstWouldAddEndPuncttrue
\mciteSetBstMidEndSepPunct{\mcitedefaultmidpunct}
{\mcitedefaultendpunct}{\mcitedefaultseppunct}\relax
\EndOfBibitem
\bibitem[Rettig \latin{et~al.}(2016)Rettig, Cort{\'e}s, Chu, Fisher, Schmitt,
  Moore, Shen, Kirchmann, Wolf, and Bovensiepen]{rettig2016persistent}
Rettig,~L.; Cort{\'e}s,~R.; Chu,~J.-H.; Fisher,~I.~R.; Schmitt,~F.;
  Moore,~R.~G.; Shen,~Z.-X.; Kirchmann,~P.~S.; Wolf,~M.; Bovensiepen,~U.
  {Persistent order due to transiently enhanced nesting in an electronically
  excited charge density wave}. \emph{Nat.\ Commun.} \textbf{2016}, \emph{7},
  1--6\relax
\mciteBstWouldAddEndPuncttrue
\mciteSetBstMidEndSepPunct{\mcitedefaultmidpunct}
{\mcitedefaultendpunct}{\mcitedefaultseppunct}\relax
\EndOfBibitem
\bibitem[Chen \latin{et~al.}(2019)Chen, Wang, Wu, Ma, Wen, Wu, Li, Zhao, Wang,
  Zhang, Huang, Li, and Huang]{chen2019raman}
Chen,~Y.; Wang,~P.; Wu,~M.; Ma,~J.; Wen,~S.; Wu,~X.; Li,~G.; Zhao,~Y.;
  Wang,~K.; Zhang,~L. \latin{et~al.}  {Raman spectra and dimensional effect on
  the charge density wave transition in GdTe$_{3}$}. \emph{Appl.\ Phys.\ Lett.}
  \textbf{2019}, \emph{115}, 151905\relax
\mciteBstWouldAddEndPuncttrue
\mciteSetBstMidEndSepPunct{\mcitedefaultmidpunct}
{\mcitedefaultendpunct}{\mcitedefaultseppunct}\relax
\EndOfBibitem
\bibitem[Zong \latin{et~al.}(2019)Zong, Dolgirev, Kogar, Erge{\c{c}}en, Yilmaz,
  Bie, Rohwer, Tung, Straquadine, Wang, Yang, Shen, Li, Yang, Park, Hoffmann,
  Ofori-Okai, Kozina, Wen, Wang, Fisher, Jarillo-Herrero, and
  Gedik]{zong2019dynamical}
Zong,~A.; Dolgirev,~P.~E.; Kogar,~A.; Erge{\c{c}}en,~E.; Yilmaz,~M.~B.;
  Bie,~Y.-Q.; Rohwer,~T.; Tung,~I.-C.; Straquadine,~J.; Wang,~X. \latin{et~al.}
   {Dynamical slowing-down in an ultrafast photoinduced phase transition}.
  \emph{Phys.\ Rev.\ Lett.} \textbf{2019}, \emph{123}, 097601\relax
\mciteBstWouldAddEndPuncttrue
\mciteSetBstMidEndSepPunct{\mcitedefaultmidpunct}
{\mcitedefaultendpunct}{\mcitedefaultseppunct}\relax
\EndOfBibitem
\bibitem[Zong \latin{et~al.}(2019)Zong, Kogar, Bie, Rohwer, Lee, Baldini,
  Erge{\c{c}}en, Yilmaz, Freelon, Sie, Zhou, Straquadine, Walmsley, Dolgirev,
  Rozhkov, Fisher, Jarillo-Herrero, Fine, and Gedik]{zong2019evidence}
Zong,~A.; Kogar,~A.; Bie,~Y.-Q.; Rohwer,~T.; Lee,~C.; Baldini,~E.;
  Erge{\c{c}}en,~E.; Yilmaz,~M.~B.; Freelon,~B.; Sie,~E.~J. \latin{et~al.}
  {Evidence for topological defects in a photoinduced phase transition}.
  \emph{Nat.\ Phys.} \textbf{2019}, \emph{15}, 27--31\relax
\mciteBstWouldAddEndPuncttrue
\mciteSetBstMidEndSepPunct{\mcitedefaultmidpunct}
{\mcitedefaultendpunct}{\mcitedefaultseppunct}\relax
\EndOfBibitem
\bibitem[Kogar \latin{et~al.}(2020)Kogar, Zong, Dolgirev, Shen, Straquadine,
  Bie, Wang, Rohwer, Tung, Yang, Li, Yang, Weathersby, Park, Kozina, Sie, Wen,
  Jarillo-Herrero, Fisher, Wang, and Gedik]{kogar2020light}
Kogar,~A.; Zong,~A.; Dolgirev,~P.~E.; Shen,~X.; Straquadine,~J.; Bie,~Y.-Q.;
  Wang,~X.; Rohwer,~T.; Tung,~I.-C.; Yang,~Y. \latin{et~al.}  {Light-induced
  charge density wave in LaTe$_3$}. \emph{Nat.\ Phys.} \textbf{2020},
  \emph{16}, 159--163\relax
\mciteBstWouldAddEndPuncttrue
\mciteSetBstMidEndSepPunct{\mcitedefaultmidpunct}
{\mcitedefaultendpunct}{\mcitedefaultseppunct}\relax
\EndOfBibitem
\bibitem[Wang \latin{et~al.}(2022)Wang, Petrides, McNamara, Hosen, Lei, Wu,
  Hart, Lv, Yan, Xiao, Cha, Narang, Schoop, and Burch]{wang2022axial}
Wang,~Y.; Petrides,~I.; McNamara,~G.; Hosen,~M.~M.; Lei,~S.; Wu,~Y.-C.;
  Hart,~J.~L.; Lv,~H.; Yan,~J.; Xiao,~D. \latin{et~al.}  {Axial Higgs mode
  detected by quantum pathway interference in RTe$_{3}$}. \emph{Nature}
  \textbf{2022}, \emph{606}\relax
\mciteBstWouldAddEndPuncttrue
\mciteSetBstMidEndSepPunct{\mcitedefaultmidpunct}
{\mcitedefaultendpunct}{\mcitedefaultseppunct}\relax
\EndOfBibitem
\bibitem[Malliakas and Kanatzidis(2006)Malliakas, and
  Kanatzidis]{malliakas2006divergence}
Malliakas,~C.~D.; Kanatzidis,~M.~G. {Divergence in the behavior of the charge
  density wave in RETe$_{3}$ (RE= Rare-Earth element) with temperature and RE
  element}. \emph{J.\ Am.\ Chem.\ Soc.} \textbf{2006}, \emph{128},
  12612--12613\relax
\mciteBstWouldAddEndPuncttrue
\mciteSetBstMidEndSepPunct{\mcitedefaultmidpunct}
{\mcitedefaultendpunct}{\mcitedefaultseppunct}\relax
\EndOfBibitem
\bibitem[Ru \latin{et~al.}(2008)Ru, Condron, Margulis, Shin, Laverock, Dugdale,
  Toney, and Fisher]{ru2008effect}
Ru,~N.; Condron,~C.~L.; Margulis,~G.~Y.; Shin,~K.~Y.; Laverock,~J.;
  Dugdale,~S.~B.; Toney,~M.~F.; Fisher,~I.~R. {Effect of chemical pressure on
  the charge density wave transition in rare-earth tritellurides RTe$_{3}$}.
  \emph{Phys.\ Rev.\ B} \textbf{2008}, \emph{77}, 035114\relax
\mciteBstWouldAddEndPuncttrue
\mciteSetBstMidEndSepPunct{\mcitedefaultmidpunct}
{\mcitedefaultendpunct}{\mcitedefaultseppunct}\relax
\EndOfBibitem
\bibitem[Banerjee \latin{et~al.}(2013)Banerjee, Feng, Silevitch, Wang, Lang,
  Kuo, Fisher, and Rosenbaum]{banerjee2013charge}
Banerjee,~A.; Feng,~Y.; Silevitch,~D.~M.; Wang,~J.; Lang,~J.~C.; Kuo,~H.-H.;
  Fisher,~I.~R.; Rosenbaum,~T.~F. {Charge transfer and multiple density waves
  in the rare earth tellurides}. \emph{Phys.\ Rev.\ B} \textbf{2013},
  \emph{87}, 155131\relax
\mciteBstWouldAddEndPuncttrue
\mciteSetBstMidEndSepPunct{\mcitedefaultmidpunct}
{\mcitedefaultendpunct}{\mcitedefaultseppunct}\relax
\EndOfBibitem
\bibitem[Montanaro \latin{et~al.}(2022)Montanaro, Giusti, Zanfrognini,
  Di~Pietro, Glerean, Jarc, Rigoni, Mathengattil, Varsano, Rontani, Perucchi,
  Molinari, and Fausti]{montanaro2022anomalous}
Montanaro,~A.; Giusti,~F.; Zanfrognini,~M.; Di~Pietro,~P.; Glerean,~F.;
  Jarc,~G.; Rigoni,~E.~M.; Mathengattil,~S.~Y.; Varsano,~D.; Rontani,~M.
  \latin{et~al.}  Anomalous non-equilibrium response in black phosphorus to
  sub-gap mid-infrared excitation. \emph{Nat.\ Commun.} \textbf{2022},
  \emph{13}, 1--7\relax
\mciteBstWouldAddEndPuncttrue
\mciteSetBstMidEndSepPunct{\mcitedefaultmidpunct}
{\mcitedefaultendpunct}{\mcitedefaultseppunct}\relax
\EndOfBibitem
\bibitem[Kenji \latin{et~al.}(1998)Kenji, Hase, Harima, Nakashima, Tani, Sakai,
  Negishi, and Inoue]{kenji1998coherent}
Kenji,~K.; Hase,~M.; Harima,~H.; Nakashima,~S.; Tani,~M.; Sakai,~K.;
  Negishi,~H.; Inoue,~M. {Coherent phonons from the CDW state in
  $\eta$-Mo$_{4}$O$_{11}$}. \emph{Phys.\ Rev.\ B} \textbf{1998}, \emph{58},
  R7484\relax
\mciteBstWouldAddEndPuncttrue
\mciteSetBstMidEndSepPunct{\mcitedefaultmidpunct}
{\mcitedefaultendpunct}{\mcitedefaultseppunct}\relax
\EndOfBibitem
\bibitem[Rettig \latin{et~al.}(2014)Rettig, Chu, Fisher, Bovensiepen, and
  Wolf]{rettig2014coherent}
Rettig,~L.; Chu,~J.-H.; Fisher,~I.~R.; Bovensiepen,~U.; Wolf,~M. Coherent
  dynamics of the charge density wave gap in tritellurides. \emph{Faraday
  Discuss.} \textbf{2014}, \emph{171}, 299--310\relax
\mciteBstWouldAddEndPuncttrue
\mciteSetBstMidEndSepPunct{\mcitedefaultmidpunct}
{\mcitedefaultendpunct}{\mcitedefaultseppunct}\relax
\EndOfBibitem
\bibitem[Lei \latin{et~al.}(2019)Lei, Duppel, Lippmann, Nuss, Lotsch, and
  Schoop]{lei2019charge}
Lei,~S.; Duppel,~V.; Lippmann,~J.~M.; Nuss,~J.; Lotsch,~B.~V.; Schoop,~L.~M.
  {Charge density waves and magnetism in topological semimetal candidates
  GdSbxTe$_{2-x-\delta}$}. \emph{Adv.\ Quantum Technol.} \textbf{2019},
  \emph{2}, 1900045\relax
\mciteBstWouldAddEndPuncttrue
\mciteSetBstMidEndSepPunct{\mcitedefaultmidpunct}
{\mcitedefaultendpunct}{\mcitedefaultseppunct}\relax
\EndOfBibitem
\bibitem[Lei \latin{et~al.}(2021)Lei, Teicher, Topp, Cai, Lin, Cheng, Salters,
  Rodolakis, McChesney, Lapidus, Yao, Krivenkov, Marchenko, Varykhalov, Ast,
  Car, Cano, Vergniory, Ong, and Schoop]{lei2021band}
Lei,~S.; Teicher,~S. M.~L.; Topp,~A.; Cai,~K.; Lin,~J.; Cheng,~G.;
  Salters,~T.~H.; Rodolakis,~F.; McChesney,~J.~L.; Lapidus,~S. \latin{et~al.}
  {Band engineering of Dirac semimetals using charge density waves}.
  \emph{Adv.\ Mater.} \textbf{2021}, 2101591\relax
\mciteBstWouldAddEndPuncttrue
\mciteSetBstMidEndSepPunct{\mcitedefaultmidpunct}
{\mcitedefaultendpunct}{\mcitedefaultseppunct}\relax
\EndOfBibitem
\bibitem[Montanaro \latin{et~al.}(2020)Montanaro, Giusti, Colja, Brajnik,
  Marciniak, Sergo, De~Angelis, Glerean, Sparapassi, Jarc, Carrato, Cautero,
  and Fausti]{montanaro2020visible}
Montanaro,~A.; Giusti,~F.; Colja,~M.; Brajnik,~G.; Marciniak,~A. M.~A.;
  Sergo,~R.; De~Angelis,~D.; Glerean,~F.; Sparapassi,~G.; Jarc,~G.
  \latin{et~al.}  {Visible pump--mid infrared pump--broadband probe:
  Development and characterization of a three-pulse setup for single-shot
  ultrafast spectroscopy at 50 kHz}. \emph{Rev.\ Sci.\ Instrum.} \textbf{2020},
  \emph{91}, 073106\relax
\mciteBstWouldAddEndPuncttrue
\mciteSetBstMidEndSepPunct{\mcitedefaultmidpunct}
{\mcitedefaultendpunct}{\mcitedefaultseppunct}\relax
\EndOfBibitem
\bibitem[Kirby \latin{et~al.}(2020)Kirby, Ferrenti, Weinberg, Klemenz, Oudah,
  Lei, Weber, Fausti, Scholes, and Schoop]{kirby2020}
Kirby,~R.~J.; Ferrenti,~A.; Weinberg,~C.; Klemenz,~S.; Oudah,~M.; Lei,~S.;
  Weber,~C.~P.; Fausti,~D.; Scholes,~G.~D.; Schoop,~L.~M. {Transient Drude
  response dominates near-infrared pump-probe reflectivity in nodal-line
  semimetals ZrSiS and ZrSiSe}. \emph{J.\ Phys.\ Chem.\ Lett.} \textbf{2020},
  \emph{11}, 6105--6111\relax
\mciteBstWouldAddEndPuncttrue
\mciteSetBstMidEndSepPunct{\mcitedefaultmidpunct}
{\mcitedefaultendpunct}{\mcitedefaultseppunct}\relax
\EndOfBibitem
\bibitem[Lei \latin{et~al.}(2020)Lei, Lin, Jia, Gray, Topp, Farahi, Klemenz,
  Gao, Rodolakis, McChesney, Ast, Yazdani, Burch, Wu, Ong, and
  Schoop]{lei2020high}
Lei,~S.; Lin,~J.; Jia,~Y.; Gray,~M.; Topp,~A.; Farahi,~G.; Klemenz,~S.;
  Gao,~T.; Rodolakis,~F.; McChesney,~J.~L. \latin{et~al.}  {High mobility in a
  van der Waals layered antiferromagnetic metal}. \emph{Sci.\ Adv.}
  \textbf{2020}, \emph{6}, eaay6407\relax
\mciteBstWouldAddEndPuncttrue
\mciteSetBstMidEndSepPunct{\mcitedefaultmidpunct}
{\mcitedefaultendpunct}{\mcitedefaultseppunct}\relax
\EndOfBibitem
\end{mcitethebibliography}

\providecommand{\noopsort}[1]{}\providecommand{\singleletter}[1]{#1}%
\providecommand{\latin}[1]{#1}
\makeatletter
\providecommand{\doi}
  {\begingroup\let\do\@makeother\dospecials
  \catcode`\{=1 \catcode`\}=2 \doi@aux}
\providecommand{\doi@aux}[1]{\endgroup\texttt{#1}}
\makeatother
\providecommand*\mcitethebibliography{\thebibliography}
\csname @ifundefined\endcsname{endmcitethebibliography}
  {\let\endmcitethebibliography\endthebibliography}{}

\end{document}